\documentclass[journal=ancac3,manuscript=article]{achemso}

\usepackage{achemso} 

\usepackage[version=3]{mhchem} 
\usepackage[T1]{fontenc}       

\usepackage{graphicx}
\usepackage{amssymb}
\usepackage{amsmath}
\usepackage{txfonts}
\usepackage{bm}
\usepackage{float}

\newcommand*\rfrac[2]{{}^{#1}\!/_{#2}}

\title{Strong Exciton-Photon Coupling in a Nanographene Filled Microcavity}

\author{David M Coles}
\affiliation{Department of Physics \& Astronomy, University of Sheffield, Sheffield S3 7RH, UK}
\altaffiliation{Contributed equally to this work}

\author{Qiang Chen}
\affiliation{Max Planck Institute for Polymer Research, Mainz D-55128, Germany}
\altaffiliation{Contributed equally to this work}

\author{Lucas C Flatten}
\affiliation{Department of Materials, University of Oxford, Oxford OX1 3PH, UK}

\author{Jason M Smith}
\affiliation{Department of Materials, University of Oxford, Oxford OX1 3PH, UK}

\author{Klaus M\"{u}llen}
\email{muellen@mpip-mainz.mpg.de}
\affiliation{Max Planck Institute for Polymer Research, Mainz D-55128, Germany}

\author{Akimitsu Narita}
\email{narita@mpip-mainz.mpg.de}
\affiliation{Max Planck Institute for Polymer Research, Mainz D-55128, Germany}

\author{David G Lidzey}
\email{d.g.lidzey@sheffield.ac.uk}
\affiliation{Department of Physics \& Astronomy, University of Sheffield, Sheffield S3 7RH, UK}


\begin{document}

\begin{abstract}
Dibenzo[\emph{hi,st}]ovalene (DBOV) - a quasi-zero-dimensional `nanographene' - displays strong, narrow, and well-defined optical-absorption transitions at room temperature. On placing a DBOV-doped polymer film into an optical microcavity, we demonstrate strong coupling of the \textbf{0 $\rightarrow$ 0'} electronic and \textbf{0 $\rightarrow$ 1'} vibrational transitions to a confined cavity mode, with coupling energies of 104 meV and 40 meV, respectively. Photoluminescence measurements indicate that the polariton population is distributed between the lower and middle polariton branches at energies approximately coincident with the emission of the DBOV, indicating polariton population via an optical pumping mechanism.
\end{abstract}

Graphene has fascinating electronic properties as demonstrated by its high charge-carrier mobility, however it does not possess a band-gap; a feature that limits its applications in transistors and optical devices\cite{ novoselov_a_roadmap_for}. A number of approaches have been explored to overcome this obstacle, including doping\cite{ohta_controlling_the_electronic,castro_biased_bilayer_graphene}, controlled interactions with a substrate\cite{balog_bandgap_opening_in,nevius_semiconducting_graphene} and the application of electric fields to graphene bilayers\cite{zhang_direct_observation_of}. However the structural confinement of the graphene sheet into a few nanometers appears to a promising strategy to create a band-gap, with `nanographene' structures developed including quasi-one-dimensional graphene nanoribbons (GNRs)\cite{han_energy_band_gap} or quasi-zero-dimensional graphene quantum dots (GQDs)\cite{ yan_colloidal_graphene_quantum,bacon_graphene_quantum_dots}. Top-down fabrication methods do not however allow the preparation of nanographenes with well-defined structures and properties, and thus we have developed a bottom-up chemical synthesis route to prepare atomically precise GNRs and GQDs\cite{narita_new_advances_in, wu_graphenes_as_potential}. Here, key optical and electronic properties such as optical absorption and energy gaps can be efficiently controlled by modulating the size, geometry, and edge structure of the nanographenes, making them promising for (opto)electronic applications\cite{ tan_atomically_precise, osella_graphene_nanoribbons_as, chen_chemical_vapor_deposition}.
 
There is significant growing interest in the development of new materials for polariton condensation; polaritons are quasi-particles that are formed in optical cavities when an optically active material reversibly exchanges energy with a resonant electromagnetic (EM) field. Here, the energy levels of the transition dipole and EM-field become hybridized, resulting in new energy eigenstates that can be described as part excitation and part photon. A convenient method to reach the strong coupling regime is to place a semiconductor material into an optical microcavity. Providing that the interaction strength between the excitons and the confined electromagnetic field is dominant over their respective loss channels, the excitons can couple to the confined cavity mode to form exciton-polaritons. Because the cavity mode energy dispersion effectively forms a trap in momentum space, it is possible to build a large population of polaritons at the bottom of the trap with the bosonic polaritons undergoing condensation at high occupation density \cite{kasprzak_bose_einstein_condensation}. This can result in the generation of non-linear phenomena such as inversionless lasing\cite{schneider_an_electrically_pumped} and polariton superfluidity\cite{amo_collective_fluid_dynamics}.

\begin{center}
\begin{figure}[t!]
\includegraphics[width=0.7\textwidth]{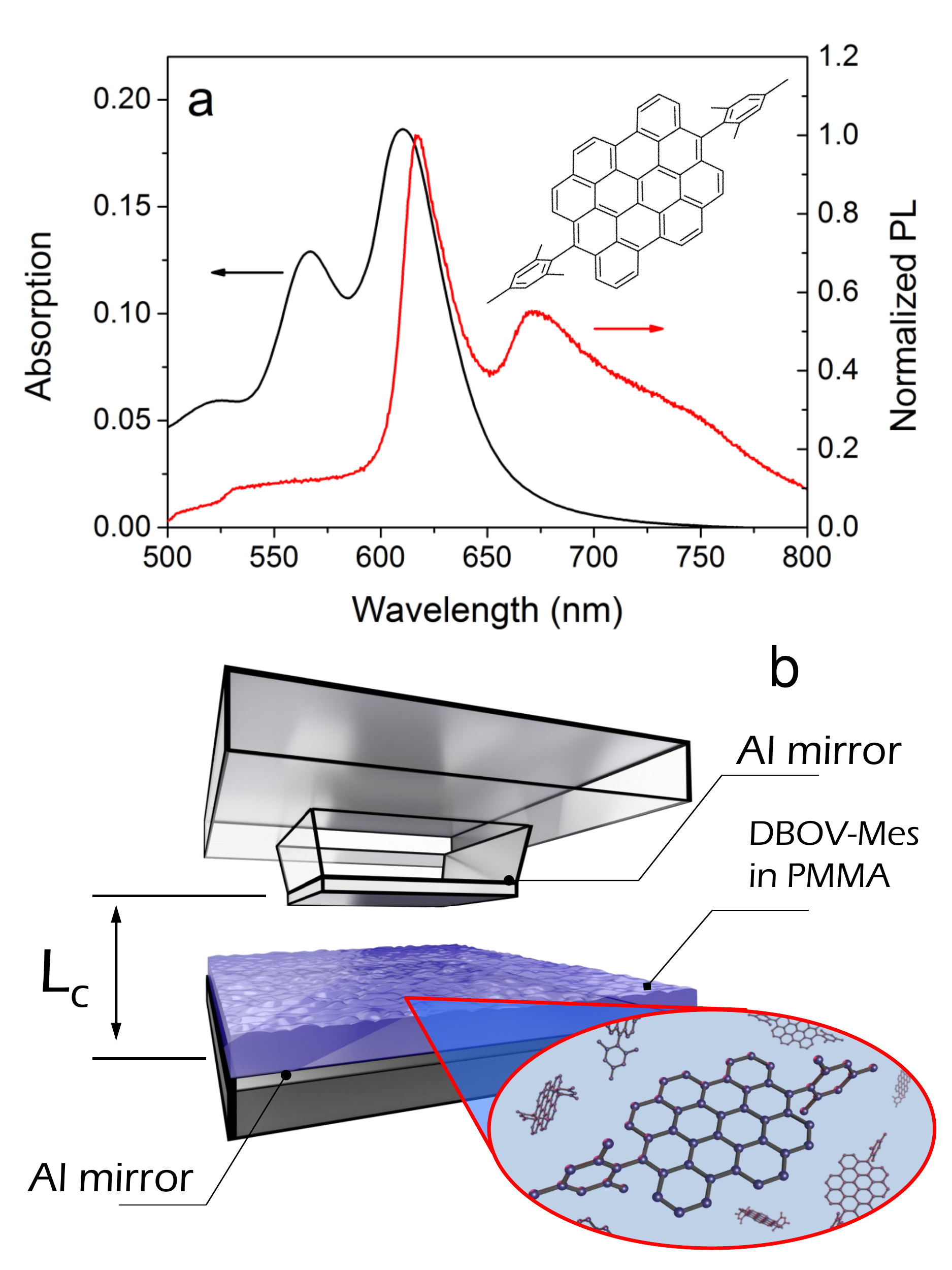}
\caption{(a) Absorption (black line) and photoluminescence (red line) of DBOV-Mes in a PMMA matrix. Left inset: Energy level diagram. Right inset: chemical structure of DBOV-Mes. (b) Microcavity structure consisting of a DBOV-Mes:PMMA film between two metallic mirrors with tunable separation, L$_{\text{c}}$.}
\label{fig1}
\end{figure}
\end{center}
 
The strong coupling regime has been demonstrated with a range of different semiconductor materials and hetrerostrucutres, including bulk semiconductors\cite{vincent_observation_of_rabi}, semiconductor quantum wells\cite{weisbuch_observation_of_the}, 2-dimensional transition metal dichalcogenides \cite{liu_strong_light_matter,flatten_room_temperature_exciton}, and organic (molecular) dyes\cite{lidzey_strong_exciton_photon,hutchinson_modifying_chemical_landscapes,ebbeson_hybrid_light_matter}. More recently, single-walled carbon nanotubes have been shown to undergo strong-coupling in a microcavity\cite{graf_near_infrared_exciton}. Nevertheless, to the best of our knowledge, such strong coupling has never been observed in graphene-based materials. Graphene in optical cavities has been previously studied \cite{engel_light_matter_interaction,furchi_microcavity_integrated}, however in those studies the materials remained firmly in the weak coupling regime whereby the cavity acts to perturb the transition rates of the graphene (the `Purcell effect'), rather than create entirely new polariton states. To explore whether nanographene materials can reach the strong coupling regime, we have explored the application of a new nanographene material (dibenzo[\emph{hi,st}]ovalene [DBOV]) in a microcavity. DBOV is particularly interesting for this application as it combines high oscillator strength and well-resolved electronic and vibrational transitions\cite{narita_dbov_synthesis}. We place this material in an open-cavity architecture and demonstrate that strong-coupling occurs to both the electronic and vibrational states of DBOV, with polaritonic emission evidenced at room temperature. We believe that the relatively high quantum efficiency of luminescence and high chemical stability of such materials make them promising candidates for polariton condensation and use in non-linear polariton-based devices.

The DBOV derivative we have explored (DBOV-Mes) has been synthesized though a bottom-up chemical process (see SI for details). The chemical structure of DBOV-Mes is shown in the inset of Figure \ref{fig1}(a), and consists of a polycyclic aromatic hydrocarbon core with attached solubilising mesityl groups that also provide kinetic protection to the relatively reactive zigzag edges. To process DBOV-Mes, it was dissolved in a solution of poly(methyl methacrylate) (PMMA) in dichloromethane (15 mg/ml, 1:1 DBOV-Mes:PMMA weight ratio) and then spin-cast to form a 540 nm thick film. The absorption of a typical film is shown in Figure \ref{fig1}(a) (black line). Here, it can be seen that there is a strong peak in absorption at 610 nm having a full width at half maximum (FWHM) linewidth of 41 nm (135 meV) corresponding to the \textbf{0 $\rightarrow$ 0'} electronic transition, with weaker vibronic replicas at 565 nm (\textbf{0 $\rightarrow$ 1'}) with a FWHM of 36 nm (139 meV) and 518 nm (\textbf{0 $\rightarrow$ 2'}) with a FWHM of 48 nm (220 meV). Following excitation at 405 nm, the film emits strong photoluminescence as shown plotted using a red line. Here, the \textbf{0' $\rightarrow$ 0} transition is prominent, having a Stokes shift of $<$10 nm. We believe that the relatively narrow linewidth of the optical and vibrational transitions and small Stokes shift point to a highly rigid molecular structure, that results in a low degree of conformationally induced broadening. Furthermore, the DBOV-Mes sample used here consists of GQDs having a single, atomically-defined structure: a feature that prohibits inhomogeneous broadening typical in top-down-fabricated GQDs that consist of a complex mixture of undefined structures.

To observe strong coupling, the GQD/PMMA film was deposited by spin-coating on a 20 nm thick semi-transparent aluminium mirror. A second mirror that had been deposited onto a small raised plinth ($\sim$ 100 $\muup$m square) was brought into close proximity ($<$ 1 $\muup$m) using a piezoelectric actuator, giving precise control over the mirror separation through application of a voltage. This forms an open cavity configuration as shown in Figure \ref{fig1}(b). Here, the use of a plinth facilitates the close positioning of the mirrors without problems resulting from the presence of unwanted dust or surface contamination which might otherwise prevent the mirrors from `closing'. To study the optical properties of the cavity, white light was focused onto the cavity, with the transmitted light imaged onto the entrance slit of an imaging CCD spectrometer using an infinity-corrected long working distance objective lens.

\begin{center}
\begin{figure}[t!]
\includegraphics[width=0.8\textwidth]{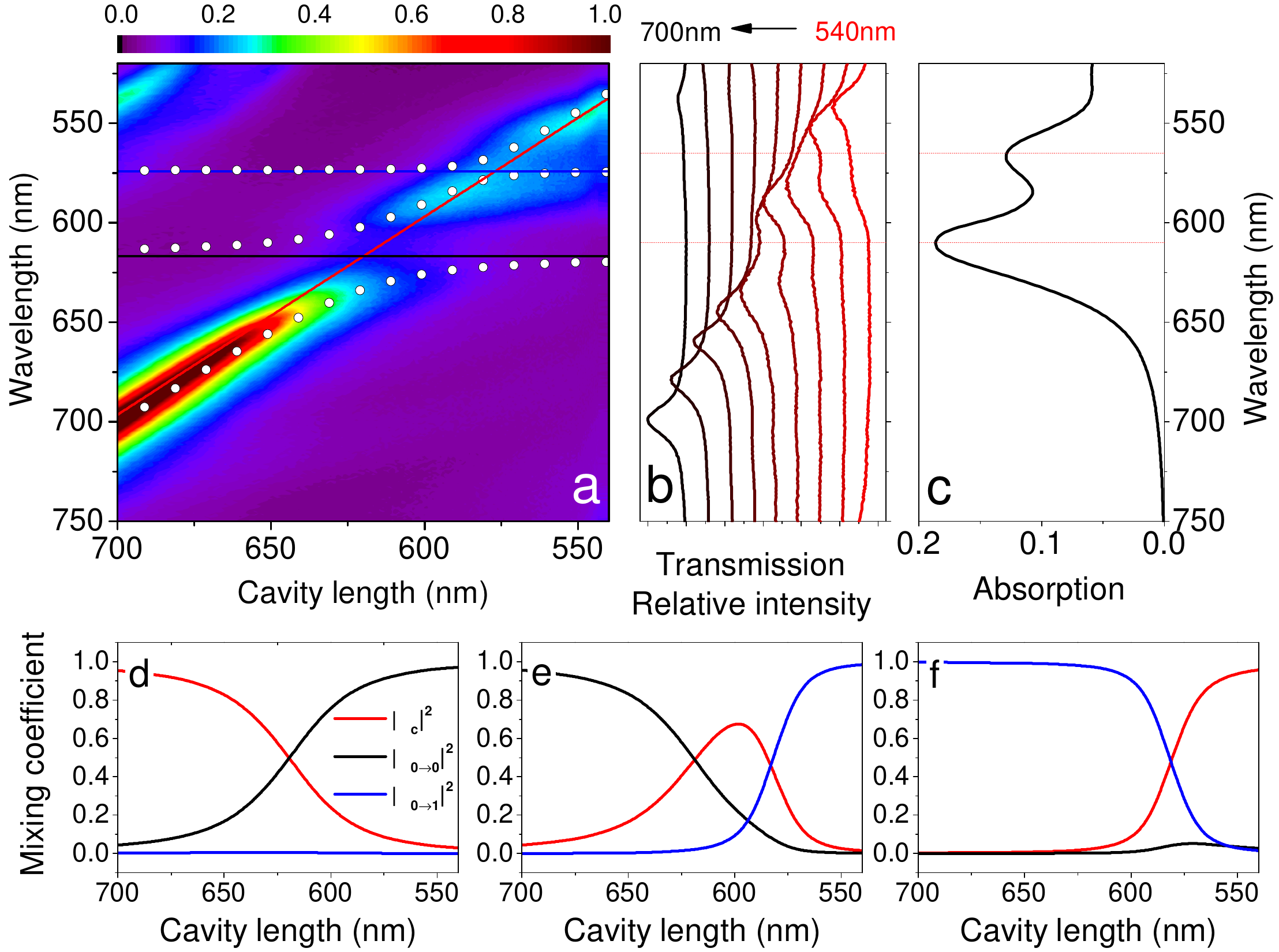}
\caption{(a) Cavity transmission spectra as a function of optical cavity length, L$_{\text{c}}$. Solid lines are fits to the data from a coupled oscillator model (black and blue lines are uncoupled exciton wavelengths, red line is uncoupled photon wavelength). Open symbols are fitted polariton branch wavelengths. (b) Normalized cavity transmission spectra for various cavity lengths. The length difference between successive plots is 20 nm. (c) Absorption spectrum of a thin film of DBOV-Mes in a PMMA marix. The polariton branch mixing coefficients are shown for (d) the lower polariton branch, (e) middle polariton branch and (f) upper polariton branch.}
\label{fig2}
\end{figure}
\end{center}

The cavity path-length, $L_{c}$ was first reduced to the order of a few microns. At this point, well defined Fabry-Perot peaks become visible in the transmission spectrum having wavelengths $\lambda_{q}=\frac{2nL_{c}}{q}$ where $q$ is the mode number and $n$ is the intracavity refractive index. The mode number for each peak can be identified from its wavelength and the wavelength of the adjacent mode on the low energy side via $q=\frac{\lambda_{q-1}}{\lambda_{q-1}-\lambda_{q}}$. The transmission through the cavity as a function of wavelength and cavity length is shown in Figure \ref{fig2}(a). Here, a peak in transmission is observed at $\sim$700 nm for a cavity length of 700 nm corresponding to the $q=3$ cavity mode. The linewidth of this peak is 19 nm (49 meV), corresponding to a $Q$ factor of $\sim$35. As the cavity length is further reduced, the mode energy increases and approaches the \textbf{0 $\rightarrow$ 0'} transition of the DBOV-Mes. At a cavity length of 620 nm, this mode splits forming polariton branches that anticross about the \textbf{0 $\rightarrow$ 0'} transition energy; a characteristic signature of strong coupling. As the mirror separation is further reduced, the mode energy increases and again splits at a cavity length of 580 nm as it undergoes anticrossing about the \textbf{0 $\rightarrow$ 1'} transition. For completeness, Figure \ref{fig2}(b) plots transmission spectra recorded from the cavity as it is closed at length intervals of 20 nm. Here the splitting of the peaks as they approach the electronic and vibrational resonances of DBOV-Mes is clearly visible.

This coupled system can be described by three coupled classical-oscillators as expressed by Equation \ref{coupled_oscillator}.

\begin{equation}
\left( \begin{array}{ccc}
E_{c} & \rfrac{\hbar\Omega_{0 \rightarrow 0'}}{2} & \rfrac{\hbar\Omega_{0 \rightarrow 1'}}{2} \\
\rfrac{\hbar\Omega_{0 \rightarrow 0'}}{2} & E_{0 \rightarrow 0'} & 0 \\
\rfrac{\hbar\Omega_{0 \rightarrow 1'}}{2} & 0 & E_{0 \rightarrow 1'} \end{array} \right) \left( \begin{array}{c}
\alpha_{c}\\
\alpha_{0 \rightarrow 0'} \\
\alpha_{0 \rightarrow 1'} \end{array} \right) = E_{\text{p}} \left( \begin{array}{c}
\alpha_{c} \\
\alpha_{0 \rightarrow 0'} \\
\alpha_{0 \rightarrow 1'} \end{array} \right)
\label{coupled_oscillator}
\end{equation}
Here $E_{c}$ is the uncoupled cavity mode energy, $E_{0 \rightarrow x}$ is the energy of the $x$th vibronic transition ($x=$ 0' or 1')  and $\hbar\Omega_{0 \rightarrow x}$ is the magnitude of the energy splitting about the $x$th transition, known as the Rabi splitting energy. The resultant eigenvalues correspond to the energy of the different polariton states created and are denoted by $E_{\text{p}}$. The eigenvectors defined through the coefficients $\alpha_{c}$ and $\alpha_{0 \rightarrow x}$ describe the relative fraction of photon and $x$th vibrational transition that is mixed into each polariton. We can fit equation \ref{coupled_oscillator} to the experimentally-observed peak positions to yield values for the Rabi splitting energies and the mixing coefficients. There are three solutions for $E_{\text{p}}$ indicating that the polariton states are distributed over 3 branches. The lower polariton branch (LPB) resides at energies lower than the $\hbar\Omega_{0 \rightarrow 0'}$ transition, the middle polariton branch (MPB) lies between the  $\hbar\Omega_{0 \rightarrow 0'}$ and  $\hbar\Omega_{0 \rightarrow 1'}$ transition energies, while the upper polariton branch (UPB) is found at energies higher than the  $\hbar\Omega_{0 \rightarrow 1'}$ transition. The fitted polariton branches are shown as white circles in Figure \ref{fig2}(a), along with the optical transition energies and cavity mode energy (black/blue lines and red line respectively). We find that the Rabi splitting energy about the electronic (\textbf{0 $\rightarrow$ 0'}) transition is $\hbar\Omega_{0 \rightarrow 0'}=$ 104 meV, while the splitting about the first vibronic replica (\textbf{0 $\rightarrow$ 1'}) is $\hbar\Omega_{0 \rightarrow 1'}=$ 40 meV. Figure \ref{fig2}(c) shows the absorption of DBOV-Mes relative to the polariton anticrossings. The definition of strong coupling requires that the sum of the cavity mode and transition half-width-at-half-maximum (HWHM) be less than the Rabi splitting energy\cite{bajoni_polariton_lasers_hybrid}. This criteria is apparently fulfilled for the coupling about the \textbf{0 $\rightarrow$ 0'} transition, however it is not reached for the \textbf{0 $\rightarrow$ 1'} transmission.

The mixing coefficients  $\alpha_{c}$, $\alpha_{0 \rightarrow 0'}$ and  $\alpha_{0 \rightarrow 1'}$ can be extracted from equation \ref{coupled_oscillator} and are plotted in Figure \ref{fig2}(d), (e) and (f) for the LPB, MPB and UPB, respectively. We find that the LPB is photon-like at longer cavity lengths, and becomes increasingly hybridized with the \textbf{0 $\rightarrow$ 0'} transition at shorter cavity lengths. The UPB is largely comprised of the \textbf{0 $\rightarrow$ 1'} transition at longer cavity lengths, becoming more photon like as the cavity length is reduced. Within the MPB, the \textbf{0 $\rightarrow$ 0'} transition is hybridized with the \textbf{0 $\rightarrow$ 1'} transition via their mutual interaction with the cavity mode\cite{holmes_strong_exciton_photon}. The maximally mixed polariton state consists of 17\% of either transition and 66\% photon.

\begin{center}
\begin{figure}[h!]
\includegraphics[width=0.8\textwidth]{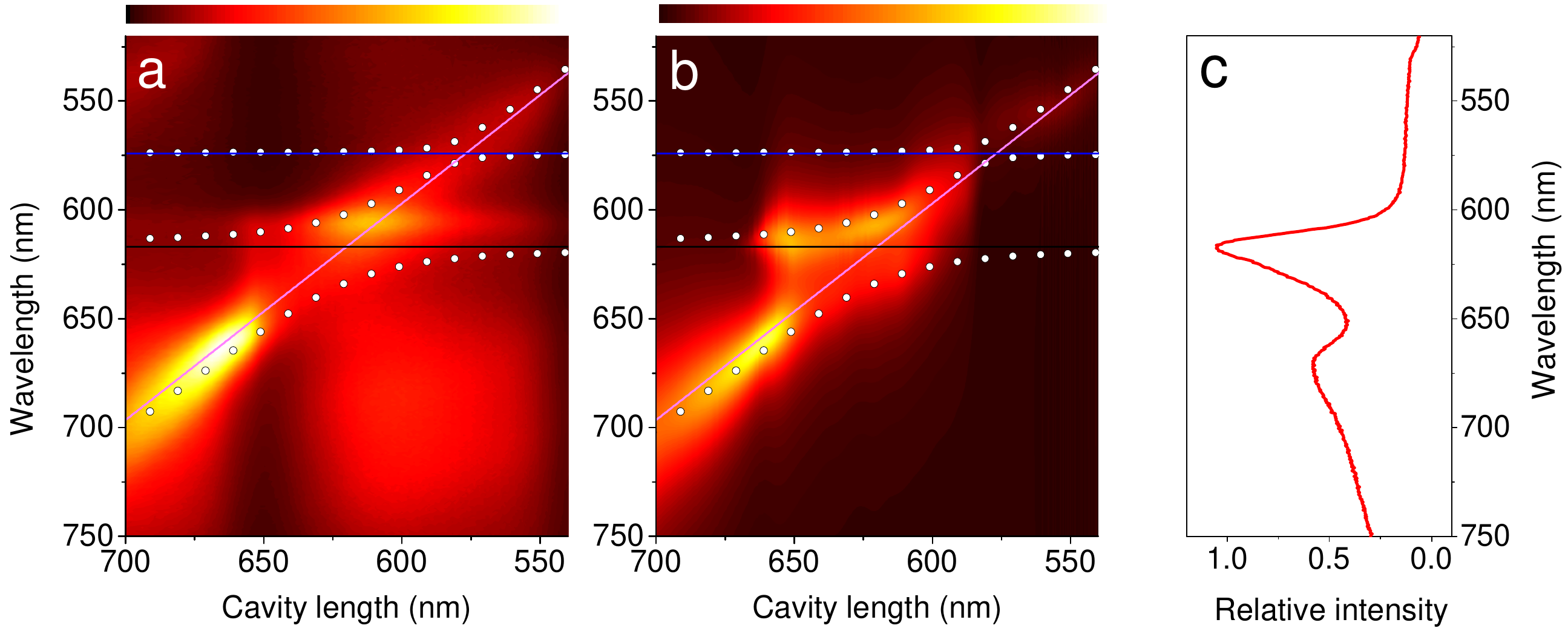}
\caption{(a) Microcavity photoluminescence as a function of optical cavity length. Lines and symbols are as in Figure 2(a). (b) Polariton population distribution. (c) PL spectrum of a thin film of DBOV-Mes in a PMMA matrix}
\label{fig3}
\end{figure}
\end{center}

We have measured the PL emission from the cavity following excitation at 405 nm as shown in Figure \ref{fig3}(a), with data plotted as a function of wavelength and cavity length. We also plot the polariton energies (as determined from transmission measurements) using circular data points. We note that the laser power admitted into the cavity is strongly modulated as the cavity length is changed, and is maximised when the excitation wavelength fulfils the cavity Bragg condition. To account for this, we record the excitation power transmitted through the cavity using a photodiode as the cavity length is scanned and use this signal to normalise the PL spectra recorded at each mirror separation. 

We find that intense PL is observed at wavelengths coincident with the LPB, with emission on the LPB peaking at 660 nm, and at 615 nm on the MPB. The polariton PL intensity ($I_{p}$) can be related to the polariton population of any particular state ($P_{p}$) through $P_{p}\propto I_{p}/\alpha_{c}$. We plot the polariton population distribution along the polariton branches in Figure \ref{fig3}(b). For comparison, we plot the free space emission of DBOV-Mes in Figure \ref{fig3}(c). Here, it can be seen that the emission originating from the LPB around 660 nm corresponds to the peak of the DBOV-Mes emission originating from the \textbf{0' $\rightarrow$ 1} transition. We also note that the peak of the polariton population along the MPB corresponds to the wavelength of \textbf{0' $\rightarrow$ 0} transition energy. This suggests that polariton states are effectively populated through an optical pumping mechanism\cite{litinskaya_fast_polariton_relaxation,michetti_exciton_phonon_scattering} in which weakly-coupled excitons in the exciton reservoir directly populate the photonic component of polariton states that are energetically degenerate with the emission. This polariton population mechanism has been shown to dominate inelastic exciton-phonon\cite{litinskaya_exciton_polaritons_in,michetti_simulation_of_j,somaschi_ultrafast_polariton_population,coles_vibrationally_assisted_polariton} or polariton-polariton scattering pathways in strongly-coupled systems based on materials having high photoluminescence quantum efficiency \cite{grant_effective_radiative_pumping}. This pumping mechanism is thought to be responsible for initiating polariton lasing in microcavities containing single anthracene crystals\cite{cohen_room_temperature_polariton}. Here the polariton population generated on the MPB results from the small Stokes shift between absorption and luminescence in this material. The peak polariton population on the LPB is 1.3 times higher than that on the MPB, despite the fact that the PL intensity of the \textbf{0' $\rightarrow$ 1} transition is almost half that of the \textbf{0' $\rightarrow$ 0} transition. Here the \textbf{0' $\rightarrow$ 1} emission transition is able to emit photons resonant with the very photon-like LPB which results in efficient polariton generation \cite{mazza_organic_based_microcavities,mazza_microscopic_theory_of}.

In summary, we have fabricated open optical microcavities containing structurally well-defined nanographene (DBOV-Mes) quantum dots. The narrow and well-separated optical transitions characteristic of DBOV-Mes lend themselves to strong exciton-photon coupling, and we demonstrate the formation of polariton states in this system at room temperature resulting from the hybridization of the cavity mode with the \textbf{0 $\rightarrow$ 0'} and \textbf{0 $\rightarrow$ 1'} vibrational transitions. Photoluminescence measurements demonstrate that polariton states in the lower and middle polariton branches are efficiently populated through an optical pumping mechanism. A large population of middle-branch polariton states is observed as a result of the small Stokes shift and high quantum yield of the \textbf{0' $\rightarrow$ 0} transition. Moreover, the high fluorescence quantum efficiency of DBOV-Mes (with values reported up to 0.79)\cite{paterno_synthesis_of} suggests that such nanographene materials may be suitable for generating polariton condensation and lasing.

\begin{acknowledgement}
We thank the UK EPSRC through Programme Grant EP/M025330/1 Hybrid Polaritonics, the Max Planck Society, and the European Commission through the Graphene Flagship for funding this research. LCF acknowledges support from the European Commission (project WASPS, 618078).
\end{acknowledgement}

\bibliography{polariton}

\end{document}